\documentclass[12pt]{article} 
\usepackage{cite}
\usepackage{latexsym}
\usepackage{amsmath}
\usepackage{pstricks}
\newcommand{\pocite}[1]{$^{(\citen{#1})}$}
\newcommand{\rcite}[1]{Ref.~\citen{#1}}

\newcommand{\becs}{\begin{cases}}
\newcommand{\bem}{\begin{matrix}}
\newcommand{\besp}{\begin{split}} 
\newcommand{\blp}{\bigl(} 
\newcommand{\brp}{\bigr)}

\newcommand{\dg}{^\circ } 
\newcommand{\encs}{\end{cases}}
\newcommand{\enm}{\end{matrix}}
\newcommand{\ensp}{\end{split}}

\newcommand{\lgl}{\langle }

\newcommand{\negg}{\sim\!} 

\newcommand{\ot}{\otimes }

\newcommand{\rgl}{\rangle }

\newcommand{\vb}{\,|\,}
\newcommand{\Tr}{{\rm Tr}}

\newcommand{\boldf}[1]{\mbox{\boldmath $#1$}} 

\newcommand{\rb}{\boldf{r}}
\newcommand{\scriptl}[1] {{\cal #1}}

\newcommand{\AS}{\scriptl A }
\newcommand{\BS}{\scriptl B }

\newcommand{\FS}{\scriptl F }

\newcommand{\PS}{\scriptl P }
\newcommand{\QS}{\scriptl Q }

\newcommand{\At}{{\tilde A}}
\newcommand{\Bt}{{\tilde B}}
\newcommand{\Ct}{{\tilde C}}
\newcommand{\Dt}{{\tilde D}}

\newcommand{\Pt}{{\tilde P}}

\newcommand{\al}{\alpha }

\newcommand{\Dl}{\Delta }

\newcommand{\om}{\omega }

\begin{document}

\title{Probabilities and Quantum Reality: Are There Correlata?}

\author{Robert B. Griffiths
\thanks{Electronic mail: rgrif@cmu.edu}\\ 
Department of Physics,
Carnegie-Mellon University,\\
Pittsburgh, PA 15213, USA}

\date{Version of 21 Sept. 2002. Submitted to Foundations of Physics}

\maketitle  

\begin{abstract}
	Any attempt to introduce probabilities into quantum mechanics faces
difficulties due to the mathematical structure of Hilbert space, as reflected
in Birkhoff and von Neumann's proposal for a quantum logic. The (consistent or
decoherent) histories solution is provided by its single framework rule, an
approach that includes conventional (Copenhagen) quantum theory as a special
case. Mermin's Ithaca interpretation addresses the same problem by defining
probabilities which make no reference to a sample space or event algebra
(``correlations without correlata'').  But this leads to severe conceptual
difficulties, which almost inevitably couple quantum theory to unresolved
problems of human consciousness.  Using histories allows a sharper quantum
description than is possible with a density matrix, suggesting that the latter
provides an ensemble rather than an irreducible single-system description as
claimed by Mermin. The histories approach satisfies the first five of Mermin's
desiderata for a good interpretation of quantum mechanics, including Einstein
locality, but the Ithaca interpretation seems to have difficulty with the first
(independence of observers) and the third (describing individual systems).
\end{abstract}


	\section{Five Desiderata Plus One}
\label{sct1}

	David Mermin is widely acknowledged to be the prince of expositors
in the field of quantum interpretation. It is a pleasure to read what he
writes, even when you don't agree with it.  When it comes to his Ithaca
interpretation of quantum mechanics\pocite{Mrmn98,Mrmn98b,Mrmn99}, I actually
agree with five out of six of Mermin's desiderata for a satisfactory
interpretation (Sec.~2 of \rcite{Mrmn98}), and before discussing some points of
serious disagreement, let me indicate the areas of overlap. Here are the five
desiderata in Mermin's own words, followed in each case by a selection from his
explanatory comments:

\begin{quote}
	1. \emph{The theory should describe an objective reality independent of
observers and their knowledge}

	\dots A satisfactory interpretation should be unambiguous about what
has objective reality and what does not, and what is objectively real should be
cleanly separated from what is ``known''.  Indeed, knowledge should not enter
at a fundamental level at all.

	2. \emph{The concept of measurement should play no fundamental role}

	There is a world out there, whether or not we choose to poke at it, and
it ought to be possible to make unambiguous statements about the character of
that world that make no reference to such probes.  A satisfactory
interpretation of quantum mechanics ought to make it clear why ``measurement''
keeps getting in the way of straight talk about the natural world;
``measurement'' ought not to be a part of that straight talk. Measurement
should acquire meaning from the theory --- not vice versa\dots  Physics ought
to describe the unobserved unprepared world.  ``We'' shouldn't have to be there
at all.

	3. \emph{The theory should describe individual systems --- not just
ensembles}

	The theory should describe individual systems because the world
contains individual systems\dots and the theory ought to describe the world and
its subsystems\dots In a nondeterministic world probability has nothing to do
with incomplete knowledge, and ought not to require an ensemble of systems for
its interpretation\dots The fact that physics cannot make deterministic
\emph{predictions} about individual systems does not excuse us from pursuing the
goal of being able to \emph{describe} them as they currently are. 

	4. \emph{The theory should describe small isolated systems without
having to invoke interactions with anything external}

	\dots I would like to have a quantum mechanics that does not require
the existence of a ``classical domain''.  Nor should it rely on quantum
gravity, or radiation escaping to infinity, or interactions with an external
environment for its \emph{conceptual} validity\dots It ought to be possible
to deal with high precision and no conceptual murkiness with small parts of the
universe if they are to high precision isolated from the rest.

	5. \emph{Objectively real internal properties of an isolated individual
system should not change when something is done to another non-interacting
system}

	\dots Einstein used [this] supposition, together with his intuitions
about what constituted a real factual situation, to conclude that quantum
mechanics offers an incomplete description of physical reality.  I propose to
explore the converse approach: assume that quantum mechanics does provide a
complete description of physical reality, insist on generalized
Einstein-locality, and see how this constrains what can be considered
physically real.
\end{quote}

	On these five desiderata I agree two hundred percent with Mermin. In
some cases I may be giving his words a slightly different interpretation from
what he intended, but at least in broad outline and probably in most of the
details, I could not agree with him more, even though I could not possibly have
expressed it with such clarity and enthusiasm.  But now we come to the sixth
desideratum:

\begin{quote}

	6. \emph{It suffices (for now) to base the interpretation of quantum
mechanics on the (yet to be supplied) interpretation of objective probability}

	I am willing at least provisionally to base an interpretation of
quantum mechanics on primitive intuitions about the meaning of probability in
individual systems. Quantum mechanics has taught us that probability is more
than just a way of dealing systematically with out own ignorance, but a
fundamental feature of the physical world.  But we do not yet understand
objective probability\dots I maintain that if we can make sense of quantum
mechanics conditional upon making sense of probability as an objective property
of an individual system, then we will have got somewhere\dots
\end{quote}

	The tone is different, and the bold confidence which characterized
desiderata 1 to 5 has changed into something more tentative.  Mermin is not
sure that this is the right way to go, and I have even greater misgivings.
Part of the problem is that I have had a great deal of difficulty making sense
of what he means by probability as an objective property of an individual
system.  While I enjoyed reading the prose, distilling the essential idea out
of Refs.~\citen{Mrmn98} and \citen{Mrmn98b} was a chore.  The main sticking
point was the notion that there can be statistical correlations among the
properties of subsystems even if those properties have no objective physical
reality: ``correlations without correlata''.  When repeated reading did not
make things clear --- something which for me is a common experience, though
quite exceptional for papers written by Mermin --- I adopted the alternative
approach of examining some of the same problems from my own point of view, in
hopes that by getting someplace using my own methods I could better make sense
of Mermin's.  The remainder of this paper is the fruit of reflections of this
sort.

	There is a fundamental difficulty when one attempts to introduce
probabilities into quantum theory by a route other than appealing to
measurement outcomes, something which both Mermin and I consider
unsatisfactory. As discussed in Sec.~\ref{sct2a}, the problem arises in trying
to relate the quantum Hilbert space, as interpreted by von
Neumann,\pocite{vNmn32t} to the sample space structure required by ordinary
probability theory.  I then discuss some ways of handling this problem,
beginning with the quantum logic approach of Birkhoff and von
Neumann\pocite{BrvN36} in Sec.~\ref{sct2b}, followed in Sec.~\ref{sct3a} by
consistent or decoherent histories --- which I simply refer to as
``histories'', since this will be unambiguous --- as developed by
Omn\`es\pocite{Omns92,Omns94,Omns99} and
me,\pocite{Grff84,Grff96,Grff98,Grff02} with major assistance from Gell-Mann
and Hartle.\pocite{GMHr90b,GMHr93} Next, in Sec.~\ref{sct3b} I comment on the
successes and limitations of the measurement approach found in standard
textbook (``Copenhagen'') quantum mechanics, as seen from a histories
perspective.
	Following this in Sec.~\ref{sct4} I return to the Ithaca
interpretation, examining in turn each of its two pillars, as defined by Mermin
in Sec.~4 of \rcite{Mrmn98}: the absence of correlata despite the existence of
correlations, which I take up in Sec.~\ref{sct4a}, and the density matrix as a
fundamental objective and irreducible property of a subsystem, the subject of
Sec.~\ref{sct4b}.  In Sec.~\ref{sct5} I respond to some comments by Mermin on
the histories approach, and then summarize my conclusions in Sec.~\ref{sct6}.

	\section{Probabilities and Logic}
\label{sct2}

	\subsection{The problem with quantum probability}
\label{sct2a}

	According to Jammer (p.~38 of \rcite{Jmmr74}), Born proposed
his probabilistic interpretation of the Schr\"odinger wave function at about
the same time that Schr\"odinger published the time-dependent version of his
equation.  The fact that two of the most important principles of
modern quantum theory did not originate in the mind of a single genius may be
one reason why they have been so hard to combine.  However, there is a much
more fundamental difficulty connected with the different mathematical
structures used for quantum mechanics and probability theory.  Quantum theory
employs a complex vector space with an inner product, a \emph{Hilbert space}.
Probability theory is founded on the notion of a \emph{sample space} (e.g.,
\rcite{Fllr68}) of mutually exclusive possibilities, one and only one of which
occurs, or is true, at any given time, or in any given experimental run.
Appropriate subsets of the sample space make up an \emph{event algebra} whose
members are assigned probabilities. For example, if one rolls a die the sample
space consists of the six possible outcomes, and if one of these, say 5 spots,
occurs, the others do not occur.  The event that the number of spots is even
belongs to the event algebra, and is assigned a probability of 1/2 for an
honest die.

\begin{figure}[h]
$$
\begin{pspicture}(-1.5,-1.5)(2.5,1.5) 
\def\lwd{0.035} 
\psset{
labelsep=2.0,
arrowsize=0.150 1,linewidth=\lwd}
\pscircle(0,0){1.5} 
\psframe(0.5,-1.0)(2.5,1.0)
\rput[l](-1.35,0){$\PS$}
\rput[r](2.35,0){$\QS$}
\end{pspicture}
$$
\caption{%
The set of points $\PS$ inside the circle correspond to the proposition $P$,
the set $\QS$ inside the square correspond to $Q$, the totality of points in
the two regions corresponds to $P\lor Q$, and the region of overlap to $P\land
Q$.}
\label{fgr1}
\end{figure}
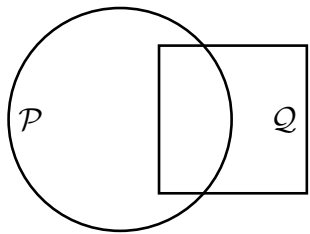

	In classical statistical mechanics the sample space is the \emph{phase
space} of classical Hamiltonian dynamics, spanned by the coordinates and momenta
of the different particles which make up the system.  If the mechanical system
is correctly represented by one of these points it is not represented by any of
the others, so one has a set of mutually-exclusive possibilities. Probabilities
are then assigned to those subsets of points in the phase space that make up
the event algebra, in a way which satisfies the rules given in texts on
probability theory.  It will be convenient to refer to such subsets as
\emph{properties}. For example the property $P$ that the energy is between 1
and 2 mJ corresponds to the subset of points $\PS$ in the phase space for which
$P$ is true.  Its negation $\Pt$, NOT $P$, corresponds to the complement
$\negg\PS$ of the set $\PS$ in the phase space, the points for which $P$ is
false (i.e., the energy lies outside the specified range). Given another
property $Q$ corresponding to a different set $\QS$ of points in the phase
space, the combined properties $P$ AND $Q$, written $P\land Q$, and $P$ OR $Q$,
written $P\lor Q$, correspond to sets of points in the phase space which can be
represented schematically in a Venn diagram, Fig.~\ref{fgr1}.  Thus in
classical statistical mechanics there is a natural correspondence between
propositions used to describe the system and sets of points in the phase space,
and between logical operations on the propositions and set-theoretical
operations on the corresponding sets.

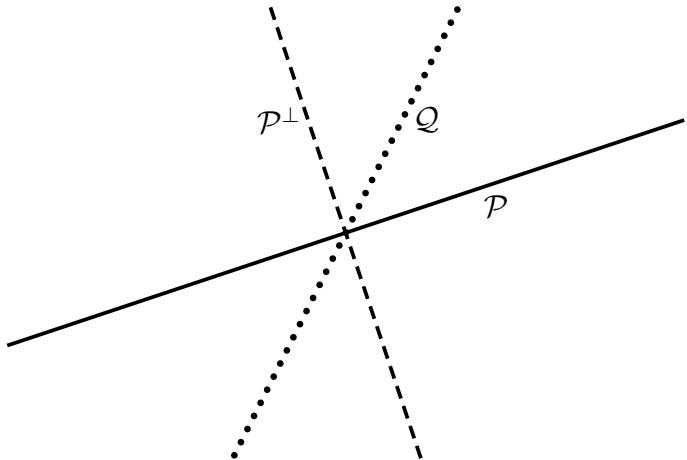
\begin{figure}[h]
$$
\begin{pspicture}(-4.5,-3.0)(4.5,3.0) 
\def\lwd{0.05} 
\psset{
labelsep=2.0,
arrowsize=0.150 1,linewidth=\lwd}
\psline(-4.5,-1.5)(4.5,1.5)
\psline[linestyle=dashed](-1,3)(1,-3)
\psline[linewidth=0.09,linestyle=dotted](-1.5,-3)(1.5,3)
\rput[t](2.0,0.5){$\PS$}
\rput[r](-0.6,1.5){$\PS^\perp$}
\rput[l](0.9,1.5){$\QS$}
\end{pspicture}
$$
\caption{%
Two-dimensional Hilbert space represented schematically by the real plane,
with origin at the point where the rays intersect.}
\label{fgr2}
\end{figure}

	The reason for mentioning such elementary matters is that in quantum
mechanics the situation is very different if we interpret the quantum Hilbert
space in the manner indicated by von Neumann in Sec.~5 of Ch.~III of
\rcite{vNmn32t}.  He associates propositions not with arbitrary collections of
points, but instead with linear subspaces of the Hilbert space. (To be precise,
with closed linear subspaces; hereafter all such qualifications are left to the
reader sophisticated enough to know they are needed.)  The simplest nontrivial
linear subspace is a \emph{ray} consisting of all kets of the form
$\{\al|\psi\rgl\}$, where $|\psi\rgl$ is some fixed vector in the space, and
$\al$ is any complex number.  One can think of this geometrically as an
infinite line through the origin, as in Fig.~\ref{fgr2} where a two-dimensional
complex Hilbert space is represented schematically by its real counterpart.
The point is that $\al|\psi\rgl$ for $\al\neq 0$ has precisely the same
physical significance as $|\psi\rgl$; the customary normalization of kets so
that $\lgl\psi|\psi\rgl=1$ is convenient but not essential, and the
arbitrariness of the phase is simply the arbitrariness of $\al$ when $|\al|=1$.
	The logical proposition $P$ associated with a ray $\al|\psi\rgl$ is
just the assertion that ``the system is in the state $|\psi\rgl$'', whatever
that may mean. For example, if $|\psi\rgl$ is a nondegenerate energy eigenstate
with energy $E$, we can interpret $P$ as ``the energy has the value $E$''.  As
an example of a larger subspace, consider the proposition that the energy of a
harmonic oscillator is less than $3\hbar\om$.  This corresponds to a
three-dimensional subspace of the Hilbert space consisting of all linear
combinations of the kets $|0\rgl$, $|1\rgl$, and $|2\rgl$, where $|n\rgl$ is
the eigenstate with energy $(n+1/2)\hbar\om$.

	The differences between a quantum Hilbert space and a classical phase
space begin to appear when one considers the negation $\Pt$ of a proposition
$P$ associated with a subspace $\PS$.  If we follow von Neumann, $\Pt$ does not
correspond to the complement $\negg\PS$ of $\PS$, the set of all vectors in the
Hilbert space which do not belong to the subspace $\PS$, but rather to the
\emph{orthogonal complement} $\PS^\perp$ of $\PS$, the collection of all
vectors $|\phi\rgl$ with the property that $|\phi\rgl$ is orthogonal to every
$|\psi\rgl$ in $\PS$ in the sense that $\lgl\phi|\psi\rgl$=0.  There are
various reasons why this is a sensible proposal.  To begin with, $\PS^\perp$ is
a subspace of the Hilbert space, whereas $\negg\PS$ is not, so that if we want
the negation of a property to itself be a property associated with a subspace,
we cannot use $\negg\PS$. Second, what distinguishes a Hilbert space from any
old (complex) vector space is the existence of the inner product
$\lgl\phi|\psi\rgl$, and since orthogonality is a rather natural and fruitful
concept from the mathematical point of view, one anticipates that it ought to
play a significant role in the physical interpretation.  Third, if $P$ is the
projector (orthogonal projection operator) onto $\PS$, then $\Pt=I-P$ (with $I$
the identity) is the projector onto $\PS^\perp$, and this way there is a nice
analogy between quantum projectors and indicator functions on the classical
phase space; for details, see Sec.~4.4 of \rcite{Grff02}. (Using the same
symbol $P$ for a proposition and the corresponding projector will cause no
confusion.)  The reasons given thus far appeal to mathematical elegance, and I
am sure this had something to do with von Neumann's choice.  But is it
\emph{physically} reasonable?  This can only be discussed by considering
physical examples.  For instance, if $P$ is the quantum proposition that the
energy of the oscillator is less than $3\hbar\om$, then its negation $\Pt$
corresponds to the subspace where the energy is greater than (or equal to, but
it cannot be equal to) $3\hbar\om$.  Is that reasonable?  I think so, but
(following a strategy I learned from Mermin) I invite any reader who disagrees
to come up with something better.

	Why all this discussion of negation?  Because it leads to a certain
oddity visible in Fig.~\ref{fgr2}, which shows one ray $\PS$ as a solid line
and its orthogonal complement $\PS^\perp$ as a dashed line.  But there are many
other rays, such as the dotted one labeled $\QS$, each associated with a
quantum proposition.  Both $\PS$ and $\QS$ are rays, the smallest possible
subspaces of the Hilbert space (aside from the origin, which does not represent
a physical property), and therefore they are the quantum counterparts of points
in a classical phase space.  But in a classical phase space, either two points
coincide, or each lies in the complement of the other, so that either they mean
the same thing, or else the truth of one implies the falsity of the other.  Now
$\PS$ and $\QS$ in Fig.~\ref{fgr2} clearly do not mean the same thing, but
$\QS$ does not lie in the subspace $\PS^\perp$ corresponding to the negation of
$\PS$, nor $\PS$ in the subspace corresponding to the negation of $\QS$, so the
truth of one does not imply that the other is false.  Thus if we follow von
Neumann in associating subspaces and their orthogonal complements with quantum
propositions and their negations, an oddity emerges in terms of nonorthogonal
rays, which have a nonclassical relationship to each other.  To put the matter
another way, it is only \emph{orthogonal} quantum states which are
\emph{distinct} in the same sense that two different classical states,
corresponding to different points in the classical phase space, are distinct.

	It helps to have a term to describe the relationship between rays such
as $\PS$ and $\QS$ in Fig.~\ref{fgr2}, which are neither identical nor
orthogonal, and we shall call them \emph{incompatible}.  Two rays are
incompatible if and only if the corresponding projectors do not commute with
each other (which means that incompatibility is very much a \emph{quantum}
notion).  Similarly, two arbitrary subspaces, or their projectors, or the
corresponding propositions, are incompatible if and only if the projectors do
not commute.  Otherwise they are \emph{compatible}.
	Quantum incompatibility in this technical sense should be clearly
distinguished from the relationship of being \emph{mutually exclusive}. Two
classical propositions which are mutually exclusive stand in a relationship
such that the truth of one implies the falsity of the other. In the quantum
case two subspaces or propositions or projectors are mutually exclusive if the
projectors are \emph{orthogonal} to each other in the sense that their product
(in either order; it makes no difference) is zero.

	As mentioned previously, ordinary (classical) probability theory is
based upon the notion of a sample space of mutually-exclusive events. It is
quite clear what this means in terms of subsets of the phase space: they are
mutually exclusive if they do not overlap, if they have no points in common.
But what should one do in the quantum case, where, as we have seen, there are
rays in the Hilbert space which are not identical but also not distinct, at
least in the same sense that nonidentical classical things are distinct?
Choosing a sample space is a fairly trivial matter in classical physics. (Well,
not exactly, but by now the rules have been worked out and the experts can
explain Borel sets to you.) But in quantum physics it is far from trivial,
because you have to decide what to do with incompatible subspaces.  This is a
fundamental problem facing anyone who wants to introduce probabilities into
quantum mechanics in a consistent fashion, at least if these probabilities are
to obey the usual rules of probability theory.  To be sure, that may be too
restrictive.  Perhaps what is needed in quantum theory is a new theory of
probability, with different rules.  In that case the fundamental problem is to
find new rules which are mathematically consistent and make physical sense.  In
any event, there is a nontrivial problem.

	\subsection{Quantum logic}
\label{sct2b}

	Since von Neumann was a skilled mathematician, it would be surprising
had he overlooked the odd effects of his scheme for associating quantum
propositions and their negations with subspaces of the Hilbert space.  He
didn't, and in a 1936 paper\pocite{BrvN36} --- one much less cited than the
1935 paper of Einstein, Podolsky, and Rosen,\pocite{EnPR35} though in my
opinion it is equally important --- he and Garrett Birkhoff tackled the problem
of the relationship between incompatible propositions by suggesting that their
conjunction $P$ AND $Q$, $P\land Q$, be associated with the intersection
$\PS\cap\QS$ of the corresponding subspaces of the Hilbert space.  This is a
natural choice in that the same operation, the intersection of two sets, is
appropriate for a classical phase space, and because the intersection of two
(closed) subspaces of the Hilbert space is another (closed) subspace. In the
case of compatible ($PQ=QP$) propositions this leads to no problems, but if one
also adopts it for incompatible propositions, the result is a logical
peculiarity.

	Rather than an abstract discussion, let us consider a two-dimensional
Hilbert space representing the spin angular momentum of a spin-half particle.
Let $P$ project on the ray $\PS$ in Fig.~\ref{fgr2} passing through the ket
$|z^+\rgl$, corresponding to the property that the $z$ component of angular
momentum $S_z$ is equal to $+1/2$ (in units of $\hbar$). Its negation $\Pt$ is
the property $S_z=-1/2$.  Let $Q$ project on the ray $\QS$ passing through
$|x^+\rgl$, corresponding to the proposition $S_x=+1/2$ for the $x$ component
of angular momentum, with negation $S_x=-1/2$.  In Birkhoff and von Neumann's
quantum logic $P\land Q$ corresponds to the intersection of the rays $\PS$ and
$\QS$, which is the origin or zero vector of the Hilbert space.  This
represents the proposition that is always false; it is the quantum counterpart
of the empty subset of the points of a classical phase space.

	One's initial reaction is that calling $S_z=+1/2$ AND $S_x=+1/2$ always
false represents good physics.  We know, or at least we were taught, that there
is no way in which one can simultaneously measure both $S_z$ and $S_x$ for a
spin-half particle, and if $S_z=+1/2$ AND $S_x=+1/2$ could ever be true, then
surely some clever experimentalist would have figured out a way to check it,
and by now would have received a Nobel prize.  However, there is a problem.
Logic tells us that the negation of a proposition that is always false is a
proposition that is always true.  The negation of $S_z=+1/2$ AND $S_x=+1/2$,
following the usual rules, is $S_z=-1/2$ OR $S_x=-1/2$.  Is it \emph{always}
true that either $S_z=-1/2$ or $S_x=-1/2$ or both?  Let's consider a case in
which $S_z=+1/2$, so its negation $S_z=-1/2$ is \emph{not} true.  Are we
entitled to conclude that if $S_z=+1/2$, then it is always the case that
$S_x=-1/2$?  But that contradicts the fact (at least in the Birkhoff and von
Neumann scheme) that $S_z=+1/2$ AND $S_x=-1/2$ is always false, for precisely
the same reason that $S_z=+1/2$ AND $S_x=+1/2$ is always false.

	When you reach a contradiction in this fashion, the first thing to do
is to go back and check that you have not made some silly mistake, and if
that doesn't solve the difficulty, you suspect that one of your
assumptions is wrong.  But there is a third way, and this is what Birkhoff and
von Neumann proposed: change the rules of logic!  Their paper made the quite
specific proposal that one abandon, or at least modify, the distributive 
identities, that is 
\begin{equation}
  P\land(Q\lor R) = (P\land Q)\lor(P\land R)
\label{eqn1}
\end{equation}
and its counterpart with $\land$ and $\lor$ interchanged.  This proposal and
variations on the same theme have given rise to a considerable body of
literature.  A quarter of a century ago Jammer (Ch.~8 of \rcite{Jmmr74}) said
that the Birkhoff and von Neumann proposal had given rise to a lot of
discussion among the philosophers, but very little interest among the
physicists. That situation has not changed, and I think it regrettable.  Not
because the Birkhoff and von Neumann scheme solves the conceptual problems of
quantum mechanics, but rather because --- and it is in this respect that their
work has a parallel with Einstein, Podolsky and Rosen --- it raises a
significant issue, a problem that needs to be thought about.  There are
important \emph{logical} issues lurking in the foundations of quantum
mechanics, and ignoring them, or treating them with contempt, does not make
them go away.  Indeed, it might just be the case --- let me be honest, I think
it \emph{is} the case --- that paying more attention to logical issues would
speed up the resolution of the difficulties raised by Einstein and his
colleagues.

	To be sure, our immediate concern is with probabilities, not logic.
But the structure of ordinary probability theory, in particular the sample
space and event algebra, is closely linked with propositional logic, which
is in some sense a limiting case of probability theory, with probabilities 1
and 0 the counterparts of TRUE and FALSE.  Thus any attempt to base quantum
probabilities on the quantum logic of Birkhoff and von Neumann, or the various
variants that have appeared since, will require a new set of rules.

	\section{Histories and Measurements}
\label{sct3}

	\subsection{Quantum histories}
\label{sct3a}

	One might also think of the (consistent or decoherent) histories
approach to quantum interpretation as constituting a revised logic, but if so
it is a much more conservative revision than that proposed by Birkhoff and von
Neumann.  For it requires no modification of the rules of propositional logic,
or of ordinary probability theory, \emph{provided} --- and here is where
conservatism comes in --- one strictly limits the domain of discourse, the set
of things which can sensibly be said about a quantum system, in an appropriate
way. 

	Again, the example of a spin-half particle is useful for illustrating
what this does and does not mean. At least under some circumstances it is
reasonable to say of such a particle that $S_z=+1/2$, because there is a
corresponding ray in the quantum Hilbert space, so a statement of this sort is
consistent with quantum mathematics as interpreted by von Neumann.  Similarly,
it is (sometimes) reasonable to say that $S_x=+1/2$, or that $S_w=+1/2$ where
$w$ denotes any direction in space.  On the other hand, $S_z=+1/2$ AND
$S_x=+1/2$ is a meaningless statement, for there is nothing in the Hilbert
space that could conceivably correspond to it.  Every ray signifies $S_w=+1/2$
for some direction $w$, and there are none left over to represent a proposition
with AND in it of the sort we are considering.
	At first it might seem as as if calling $S_z=+1/2$ AND $S_x=+1/2$
meaningless is the same as saying that it is always false, and in ordinary
language there is, indeed, not much difference between the two. However, the
negation of a false but meaningful statement is a true statement, whereas the
negation of a meaningless statement is equally meaningless, neither true nor
false.  The sense of ``meaningless'' employed here is the same as when in
ordinary logic one combines two meaningful propositions $P$ and $Q$ in the form
$P\land\lor Q$.  Such an expression does not conform to the rules given in
books on logic for forming meaningful sentences, and hence it is a waste of
time discussing what it means, whether it is true or false, or what its
negation might be.

	What the histories approach does is specify that a proper domain of
logical discourse in quantum theory is limited to a set of \emph{compatible}
propositions corresponding to subspaces with projectors that commute with each
other. Within such a \emph{framework} (my terminology) or \emph{logic}
(Omn\`es) all the usual rules of (classical!)  propositional logic work just
fine, without the slightest modification.  Logicians know, and the books on
logic tell us, that logical reasoning requires a well-defined domain of
discourse or collection of propositions (the well-formed sentences, or
whatever), and that specifying this collection is a task, sometimes a
nontrivial one, which needs to be done before you draw your conclusions. We
physicists, amateurs in this as in every other field outside our specialty, are
inclined to plunge ahead and leave the mathematical or logical niceties to some
future paper or (more often) to someone else.  Frequently we get away with it,
but not always, and if Omn\`es and I are right, quantum mechanics is one case
in which it pays to pay attention to what you are doing.

	In classical physics we do not have to be specific about the domain of
discourse, because in most instances only one reasonable possibility will come
to the mind of a trained physicist, so there is no point spending time at the
beginning of a conversation or a paper stating what it is.  As soon as I
mention the angular momentum of a spinning top, you know precisely what the
phase space is, so you know what constitutes a meaningful description. On the
other hand, in quantum mechanics there are always many different frameworks
(logics) to choose from.  You might be interested in talking about the $z$
component of angular momentum of a spin-half particle, in which case you use
the $S_z$ framework corresponding to the orthonormal basis $|z^+\rgl$,
$|z^-\rgl$.  These represent two mutually-exclusive possibilities, because the
product of the corresponding projectors $[z^+][z^-]$ is zero, where $[\psi]$ is
a convenient shorthand for $|\psi\rgl\lgl\psi|$, and so they constitute a
sample space suitable for a probabilistic description of the system.  On the
other hand, if you are interested in the $x$ component of the particle's spin
you use the $S_x$ framework with projectors $[x^+]$ and $[x^-]$.  The $S_z$ and
$S_x$ frameworks are \emph{incompatible}: they are logically disjoint, you
cannot combine propositions from one in a meaningful way with propositions from
another.
	
	More generally, a possible framework for discussing a quantum system at
a single time always corresponds to a \emph{decomposition of the identity}, a
collection of projectors that commute with one another and sum to the identity
$I$. They represent mutually exclusive possibilities, since the product of any
two (distinct) projectors belonging to the collection is zero.  These
projectors together with sums of two or more of them constitute the collection
of propositions that make sense in this framework.  Two frameworks are
\emph{compatible} if the projectors in one commute with the projectors in the
other; otherwise they are incompatible.  To describe a quantum system
developing in time you use a framework consisting of \emph{histories}: each
history is a sequence of projectors representing properties of the system at a
succession of times.  The rules for constructing frameworks of histories are
more complicated than those for describing a quantum system at a single time,
and in fact there has been some diversity in the proposed rules, although in
the end most of us seem to have settled on a formulation due to Gell-Mann and
Hartle.\pocite{GMHr93}
	A given framework, either for the properties of a quantum system at a
single time or for a collection of histories, is a collection of commuting
projectors, and the corresponding logical propositions relate to one another in
precisely the same way as in familiar, classical logic.  For this reason,
introducing probabilities is no problem: one simply follows the usual rules of
probability theory, and all the usual rules are satisfied, as long as the
discussion is confined to a single framework.

	But then, how are we to relate the propositions or the probabilities
that occur in different, incompatible frameworks?  To this Omn\`es and I have a
simple answer: don't!  This is stated more formally in the \emph{single
framework rule}.  It says that propositions belonging to incompatible
frameworks cannot be combined in any way, shape, or form when constructing a
meaningful description of the quantum world.
	At one level the single framework rule agrees pretty well with the
quantum physicist's intuition.  Talking about a particle which is in an energy
eigenstate of a harmonic oscillator AND has a position associated with it, or
has a position at the same time as it has a momentum, or a spin half
particle with values for both $S_x$ AND $S_z$ --- these things make us
uncomfortable.  So it is nice to have a rule which gives a reason --- the
projectors for the two items don't commute with each other --- for not doing
what would in any case make us nervous.
	But there is another level at which the rule is much less intuitive.
This is when a projector $P$ for some property that interests us belongs to
two  different, incompatible (because some \emph{other}
projectors do \emph{not} commute) frameworks $\FS_1$ and $\FS_2$.  We have just
used $\FS_1$ to compute a probability for $P$, and what could be more
reasonable than to suppose that $P$ must have the same probability in $\FS_2$?
Indeed, how could it possibly be otherwise? This sort of reasoning
leads to all manner of paradoxes, and it is by insisting on the single
framework rule that the histories approach avoids or, as I like to put it,
tames these paradoxes (see Chs.~20 to 25 of \rcite{Grff02}).

	The single framework rule has been frequently misunderstood, so it may
be worth making a couple of brief comments about what it does and does not
mean.  First, it is not a prohibition on \emph{creating} quantum descriptions
using any and every framework which strikes your fancy. You can talk about, and
it makes sense to talk about, $S_z$ or $S_x$ or whatever you please. What is
prohibited is \emph{combining} incompatible frameworks, thinking of them as
somehow both referring to the same system at the same time.  Second, the
relationship of two incompatible frameworks is \emph{not} that they are
mutually exclusive, that if one is true the other must be false.  We have
already seen that this is the wrong way (at least if you accept von Neumann's
notion of negation) of viewing two incompatible propositions, and it is equally
wrong when it comes to incompatible frameworks.  It is useless to search for
some ``law of nature'' which identifies the ``right'' framework for describing
a quantum situation, because alternative frameworks are not related in this
fashion.  I shall have more to say about this later.

	In summary, the histories approach requires reasoning in a different
way about quantum systems than we are accustomed to doing in the case of
classical systems, and in this sense requires a ``new logic''. However, what is
new is not \emph{modus ponens} or the rule of the excluded middle. Instead it
is a syntactical rule that specifies which quantum descriptions make sense, and
imposes the painful discipline (slightly less painful when you get used to it)
of paying attention to what constitutes a meaningful domain of discourse.
There are some significant benefits from exercising discipline.  One is that
you can think clearly about what is going on in a quantum system without
running into logical paradoxes. Next, quantum probabilities can be manipulated
using \emph{exactly} the same rules as for their classical counterparts. Third,
all those mysterious nonlocal influences are eliminated from quantum theory:
they owe their existence to logical errors, and correcting these errors
banishes them from the scene.

	\subsection{Quantum measurements}
\label{sct3b}

	I am in complete agreement with Mermin's desideratum 2, that the
concept of measurements should play no fundamental role in quantum
interpretation, and consequently I share his dissatisfaction with the
``Copenhagen'' interpretation, understood to mean the ideas of Bohr and the
other founding fathers as these have come down to us in the textbooks and
lectures from which we first learned the subject.  Nonetheless, one must admit
that the Copenhagen has been remarkably successful in introducing at least some
probabilities into quantum theory, namely those referring to the outcomes of
measurements, and it seems worth investigating, using various frameworks of the
sort introduced in Sec.~\ref{sct3a}, the reasons for its success, and what are
its limitations.

	One of the many possible frameworks that can be used to analyze the
time development of a quantum system is what I shall call a \emph{unitary}
framework: let $|\Psi_t\rgl$ be a solution of Schr\"odinger's equation, and at
any time let the projector $[\Psi_t]$ onto $|\Psi_t\rgl$ be part of the
decomposition of the identity used to describe what is going on.  Consider a
situation in which a measurement is about to take place, and at the initial
time $t_0$ let $|\Psi_0\rgl$ represent the ``ready'' state of the apparatus and
that of the system it will soon be measuring.  Let $|\Psi_1\rgl$ be the
resulting state at $t_1$ when the measurement interaction has taken place.  As
Schr\"odinger\pocite{Schr35t}\vphantom{\cite{WhZr83}} himself taught us, one
can easily imagine an initial $|\Psi_0\rgl$ which is physically reasonable, but
develops into a monstrosity $|\Psi_1\rgl$ that I like to call a macroscopic
quantum superposition (MQS), but everyone else refers to as ``Schr\"odinger's
cat'', because it is a superposition of states in which the apparatus pointer
indicating the outcomes has various different positions.
	The quantum historian will admit he does not understand $|\Psi_1\rgl$
any better than you do, but will add that if you insist that this represents
the physical state of the total system at time $t_1$ --- that is, if you are
using the unitary framework --- then anything else you want to say about the
system at this time must be expressed in terms of projectors which commute with
the projector $[\Psi_1]$ onto $|\Psi_1\rgl$.  In particular, you certainly
\emph{cannot} speak about a pointer pointing in a particular direction (or a
cat which is dead or alive), because whatever projector could be employed for
this purpose will not commute with $[\Psi_1]$.  The situation is analogous to
asserting that $S_z=+1/2$ and then trying to ascribe values to $S_x$. It makes
no sense.

	As seen from a histories perspective, what theorists trained in
Copenhagen actually do when they calculate measurement outcomes is to use what
I call the \emph{dragon} framework, after Wheeler's striking image of the great
smoky dragon\pocite{Whlr83} which at some instant bites the detector, at which
point the dragon collapses, to be replaced by a well-defined pointer position.
In histories language, unitary time development is employed up to just before
$t_1$, but then one introduces a physically reasonable (or at least
interpretable) \emph{pointer} decomposition of the Hilbert space identity into
subspaces (projectors) such that in each of then the pointer has a well-defined
position. Constructing such a decomposition is in principle not too difficulty,
since any two states of affairs which are distinguishable at a macroscopic
level will be represented by quantum projectors whose product is (essentially)
zero, and thus we have a good sample space of mutually exclusive possibilities.
This framework and the unitary framework constitute equally valid but mutually
incompatible ways of describing quantum time development, so it makes no sense
to combine them, and you will only confuse yourself if you think of them as
equivalent for all practical purposes or, indeed, for any purposes whatsoever.
(In saying this I am siding with Bell\pocite{Bll90} in his critique of an
earlier Ithaca interpretation, Sec.~20 of \rcite{Gttf66}.)  Note that just as
the unitary framework excludes any discussion of the probabilities of
measurement outcomes, in the same way using the dragon framework excludes any
discussion of the MQS states of different pointer positions once you have
introduced the pointer decomposition of the identity.  The precise time at
which the dragon collapses is a choice that is up the to the quantum physicist;
indeed, there are many different (incompatible) dragon frameworks, and since
there is no ``right'' framework, there is also no point in discussing when it
is that the collapse occurs. To put it another way, collapse is something that
occurs in the theorist's notebook, not the experimentalist's laboratory.

	But if collapse can be pushed around in time, is there any reason that
it cannot be pushed back to shortly \emph{before} the measurement takes place?
No there is not, and if you do this you arrive at what I shall call the
\emph{measurement} framework in which just before the measurement takes place
the soon-to-be-measured system actually has the property corresponding to the
after-the-measurement-occurs pointer position.  That means the measurement
\emph{is} actually a measurement!  One of the few idiocies to escape Bell's
scathing criticisms\pocite{Bll90} of the Copenhagen measurement approach is
that it is not a theory of \emph{measurement} in any reasonable sense of the
word.  Ordinary macroscopic measurements, such as the height of a table, tell
one a property of the measured system both before and after the measurement
takes place.  Typical measurements of properties of quantum entities in the
laboratory, such as the energy of an alpha particle, tell one what the property
was \emph{before} it was, often quite severely, altered by its interaction with
the measuring apparatus.  A quantum theory of measurement that can tell us the
probability of a pointer position, but is incapable of relating that position
to some earlier state of affairs is of no use for analyzing the experimental
data that accumulates so rapidly at particle accelerators!  Despite the
prominence it receives in textbooks, the von Neumann theory of measurement,
Ch.~VI of \rcite{vNmn32t}, in which a property of the measured system
\emph{after} the measurement takes place is correlated with the position of a
pointer indicating the outcome of the measurement, is actually of very limited
utility for discussing what actually goes on in the laboratory.  However, in
those circumstances in which it can be (properly) applied there is also a
framework which justifies the usual conclusions in terms of conditional
probabilities; for details, see Sec.~18.2 of \rcite{Grff02}.

	The measurement framework is one step in the right direction, but what
particle experimentalists actually use when designing their equipment and
analyzing their data is what I shall call the \emph{practical framework} in
which a particle \emph{described quantum mechanically} moves along what on
macroscopic length scales is a classical trajectory from the point where it was
produced (or scattered) towards whatever detector eventually detects it.  And
if it is detected at some point, you know that a short time earlier it was
headed toward that point and not in some other direction. The practical
framework, let me emphasize, provides a fully quantum-mechanical description of
what is going on, and it justifies the usual phenomenological talk in which
particles possess certain properties before a measurement takes place, or in
the complete absence of any measurement.  It also indicates the limits of
validity of such talk: why, for example, you may (but will not always) be in
trouble if your description says that the neutron passed through a definite arm
of the interferometer.  The histories approach justifies (at least most of the
time) the intuitive belief of experimental physicists that they are dealing
with real particles moving along in the manner they have been accustomed to
talking about.  For more on this topic see \rcite{Grff02c} and pp.~123ff of
\rcite{Omns94}.

	In summary, the main defect of Copenhagen, viewed from the histories
perspective, is not what it says about measurement outcomes and their
probabilities, which can be readily justified using the fully-consistent
histories interpretation, but rather its inability to push further and relate
measurement outcomes to the microscopic properties which are being measured.
In practice physicists who deal with real measurements tend to believe that
they are really measuring something: they talk as if the ``measurata'', or
whatever term Mermin approves, really exist. (For a few examples of situations
where this sort of talk is reasonable, see Sec.~3 of \rcite{Grff84}, Sec.~4.1
of \rcite{Omns94}, and Sec.~IV of \rcite{Grff02c}.)  The histories approach,
unlike Copenhagen, tells us why quantum mechanics, properly interpreted,
justifies this belief.

	One can also view the matter in a different way.  Restricting the
discussion to macroscopic measurement outcomes, which is really what the great
smoky dragon amounts to, is an example of limiting the domain of logical
discourse in a manner consistent with the single framework rule.  Sticking to
macroscopic events as they occur in the world around us is a safe thing to do
because, as seen quantum-mechanically, the macroscopic ``classical'' limit can
always be described using a single quasi-classical framework --- more on this
at the end of Sec.~\ref{sct4a}.  It is safe, but it is inadequate, for
physicists need to be able to discuss the microscopic world in a physically
meaningful way; we need to talk about particles and spins and such things.
Quantum mechanics interpreted using the single framework rule allows us to do
this without getting into contradictions.

	\section{The Ithaca Pillars}
\label{sct4}

	\subsection{Correlations without correlata}
\label{sct4a}

	The first pillar of the Ithaca interpretation is \emph{objective
probability}, the notion that correlations (more generally, probabilities) are
the only fundamental and objective properties of the world.  The term
``objective'' could mean various things. One is that quantum mechanics provides
specific rules for calculating probabilities, e.g., the probability that at the
end of the measurement the apparatus pointer will be directed at 2 rather than
1, and that two physicists who accept the principles of the theory and start
with the same initial state will assign the same values.  It is not a matter of
subjective judgments, of saying ``my hunch is the chance of its happening is
less than 1/2.''  If ``objective'' is understood in this sense, most quantum
physicists would probably say our probabilities are objective.  An alternative
interpretation could be that quantum dynamics is intrinsically stochastic
rather than deterministic, and this is an objective fact about the world.
Many, though not all, quantum physicists would agree.

	But Mermin seems to have something different in mind, since he
refers to objective probability as an \emph{irreducible} feature of the world,
not linked in any way with human ignorance.  Consider an example.  Just before
a coin is tossed I assign a probability of $1/2$ to each outcome, but after it
has been tossed and I have seen the outcome, I assign a definite value, $H$ or
$T$.  In this case the earlier uncertainty has been \emph{reduced}, the earlier
probability replaced by a new one conditioned on new data.  You can also
imagine a quantum coin toss in which $H$ and $T$ correspond to which of two
photomultipliers detects a photon which has just passed through a beam
splitter.  In any case, after the toss he who knows what happened knows more
than he who does not know.  I think Mermin intends his objective probability to
be something that cannot be reduced in this way, perhaps like a coin which is
still spinning, so it really does not make sense to say $H$ or $T$.

	In particular, Mermin asserts that \emph{correlations} between
subsystems of a quantum system are real and objective.  By this he means joint
probabilities of the form $\Pr(A,B)$, where $A$ is some property (projector)
associated with subsystem $\AS$ and $B$ a property of subsystem $\BS$, when the
total system has a Hilbert space $\AS\ot\BS$, and their generalizations to
three or more subsystems.  If $B$ is equal to the identity, $\Pr(A,B)$ becomes
the probability $\Pr(A)$ of $A$ by itself without reference to the other
subsystem $\BS$, so this is included in what Mermin refers to as a
``correlation''.  And since any system to which anyone except a cosmologist
applies quantum mechanics can be thought of as part of a larger universe, it
seems to me that Mermin is asserting the objective reality of \emph{all}
quantum probabilities of the type
\begin{equation}
  \Pr(E) = \lgl\Psi| E |\Psi\rgl,
\label{eqn2}
\end{equation}
where $E$ is any projector on the Hilbert space, and $|\Psi\rgl$ is the quantum
state of the total system. There are, to be sure, passages where he stresses
the notion of a subsystem, but I think most if not all of what he wants say
about objective probabilities can be formulated without reference to
subsystems, and is a bit clearer in this more general framework.  In further
support of this idea is the fact that Mermin does not care how one splits the
Hilbert space up into subsystems, which if it can be done at all can be done in
a large (uncountably infinite) number of ways.  Adding a bit of casuistry, any
Hilbert space is the tensor product of itself with the one-dimensional Hilbert
space of complex numbers, and in this sense it is its own subsystem.  Anyway, I
shall proceed on the assumption that the subsystem structure as such is not
central to ``correlations without correlata'', which is basically equivalent to
``probabilities without an event algebra''.

	In everyday life we do not normally think of probabilities as things
that exist, or at least their existence is less real than their
\emph{referents}, the states of affairs of which they are probabilities.
Saying the probability is 1/2 that the coin toss will result in $H$ is an
abstract notion whose significance has long been debated.  But the referent,
the actual outcome of the toss, is part of the real world, and can have
significant consequences, as when it determines which team has the ball when
the game begins.  So Mermin's including correlations, which is to say
probabilities, among real things, among what Bell would have called quantum
beables, is surprising.  But what is truly astounding is his claim that while
(at least some) quantum probabilities are real, their referents are \emph{not}
part of physical reality.  While $\Pr(A)$ (or $\Pr(A,B)$ if one considers
subsystems as essential) is an objective, real property, $A$ (or $A$ and $B$)
does not exist.  This is a truly radical approach to quantum interpretation,
and one's first inclination is to dismiss it as unintelligible nonsense.

	But Mermin is not some antiscientific crank, and other reputable
scholars have suggested that quantum theory requires a radical revision of our
view of reality, even to the point of abandoning conventional logic
(Sec.~\ref{sct2b}).  So let us try and understand what in the (quantum) world
has driven Mermin to this (seemingly) crazy conclusion.  The key passage is
Sec.~IX of \rcite{Mrmn98b} where, as in App.~B of \rcite{Mrmn98}, Mermin
discusses Hardy's paradox.\pocite{Hrdy92}  Elsewhere Mermin\pocite{Mrmn94}
gives us his own lights-and-switches version, and I strongly recommend it to
anyone who wants to get a feeling for what is genuinely paradoxical about
Hardy's paradox.  There is no need to repeat it here, for one can get to the
crucial logical and probabilistic issues without making reference to a
particular physical setup.  In a four-dimensional Hilbert space, $A$, $B$, $C$,
and $D$ are four appropriately chosen properties, each a projector onto a
2-dimensional subspace.%
\footnote{The translation to Mermin's notation in Sec.~IX of \rcite{Mrmn99}
is as follows: $A=2'G$, $\At=2'R$, $B=1R$, $\Bt=1G$, $C=1'G$, $\Ct=1'R$,
$D=2R$, $\Dt=2G$.}
 The pairs
\begin{equation}
  AC\neq CA,\quad BD\neq DB  
\label{eqn3}
\end{equation}
do not commute, whereas $A$ commutes with $B$ and $D$, as does $C$. The
negations are $\At=I-A$, $\Bt$, $\Ct$, and $\Dt$.

For a particular, carefully chosen
$|\Psi\rgl$, probabilities are assigned using \eqref{eqn2}, and one finds that
\begin{equation}
\begin{split}
  \Pr(A,\Bt) &= 0,\\
  \Pr(B,\Ct) &= 0,\\
  \Pr(C,\Dt) &= 0,\\
  \Pr(A,\Dt) &> 0.
\end{split}
\label{eqn4}
\end{equation}
Here $\Pr(A,\Bt)$ is the probability of $A$ AND $\Bt$, obtained by setting
$E=A\Bt$ in \eqref{eqn2}; $A\Bt$ is a projector because $A$ and $\Bt$ commute,
and the same is true for the other products needed in \eqref{eqn4}.

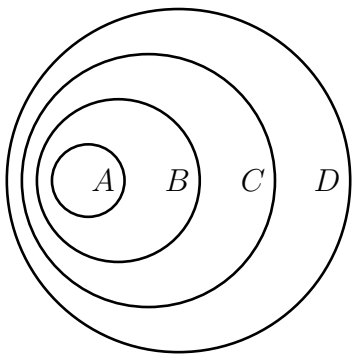
\begin{figure}[h]
$$
\begin{pspicture}(-1.1,-2.3)(3.5,2.3) 
\def\lwd{0.035} 
\psset{
labelsep=2.0,
arrowsize=0.150 1,linewidth=\lwd}
\pscircle(0.0,0){0.5}
\pscircle(0.4,0){1.1}
\pscircle(0.8,0){1.7}
\pscircle(1.2,0){2.3}
\rput[r](0.35,0){$A$}
\rput[r](1.35,0){$B$}
\rput[r](2.35,0){$C$}
\rput[r](3.35,0){$D$}
\end{pspicture}
$$
\caption{%
Venn diagram to illustrate \eqref{eqn4}.}
\label{fgr3}
\end{figure}

	There is something odd about this collection of probabilities as one
can see by looking at the Venn diagram in Fig.~\ref{fgr3} for the set of all
points in the sample space (think of it as discrete) with probability greater
than zero.  The points inside circle $A$ are all the ones (of positive
probability) for which $A$ is true, those inside $B$ (including those inside
$A$) the ones for which $B$ is true, and so forth. That $A$ lies inside $B$
follows from the fact that $\Bt$ is the set of points \emph{outside} $B$, and
the first equality in \eqref{eqn4} tells us that this set has no points in
common with $A$.  In the same way the other equalities tell us that $B$ lies
inside $C$, which lies inside $D$. Hence, as evident from the figure, $A$ has
no points in common with $\Dt$, the set of points outside $D$. But this
contradicts the final inequality in \eqref{eqn4}.

	What is wrong?  Let us first analyze the situation using the histories
approach, which demands that sample spaces be specified before probabilities
can be defined.  Since nothing has been said in advance, the sample space of
$\Pr(A,\Bt)$ must be specified implicitly by the fact that the corresponding
event algebra contains $A$ and $\Bt$.  Since $A$ commutes with $B$, and
therefore $\Bt$, there is such an event algebra, which contains negations, thus
$\At$ and $B$, and conjunctions, such as $A\land B$ corresponding to the
projector $AB$. The result is a unique sample space associated with a
particular orthonormal basis of the four-dimensional Hilbert space, with the
basis vectors representing four mutually-exclusive possibilities.  Similarly,
each of the other probabilities in \eqref{eqn4} corresponds to a particular
orthonormal basis or sample space for which the corresponding probability makes
sense.  It turns out that the four different bases associated with the four
probabilities in \eqref{eqn4} are all distinct, so each probability refers to a
different sample space.  This is also obvious from \eqref{eqn3}, since one
cannot combine sample spaces containing noncommuting projectors.  And this,
says the historian, explains what is wrong with our argument based on the Venn
diagram in Fig.~\ref{fgr3}: it assumed a \emph{single} sample space.  Hence the
contradiction was a product of careless reasoning: you cannot combine
probabilities referring to different sample spaces.

	While this noncombination rule is equally valid for classical and
quantum physics, in the former it can usually be ignored, because behind one's
probabilistic reasoning there is always a \emph{single} sample space.  Even if
it has not been explicitly identified, the fact that all probabilities refer to
this single sample space means that no contradictions of the type we are
considering can arise.  But in quantum mechanics there is not always a single
sample space. Ignoring the rules can then lead to trouble, as in the present
example, which can be thought of as a probabilistic analog of the mistake a
beginning student makes when he carelessly assumes that in quantum theory $xp$
and $px$ are the same because they are the same in classical mechanics.

	Mermin's argument follows a slightly different route.  First, using the
fact that $\Pr(A)$, $\Pr(B)$ and $\Pr(C)$ are greater than zero, he uses $
\Pr(B\vb A) = \Pr(A,B)/\Pr(A)$, etc., to rewrite \eqref{eqn4} in the equivalent
form
\begin{equation}
\begin{split}
  \Pr(B\vb A) &= 1,\\
  \Pr(C\vb B) &= 1,\\
  \Pr(D\vb C) &= 1,\\
  \Pr(D\vb A) &< 1.
\end{split}
\label{eqn5}
\end{equation}
Logically combining the first three equalities leads to the conclusion that
$\Pr(D\vb A)=1$, which contradicts the final inequality. From the fact that he
deduced a contradiction using conditional probabilities, Mermin draws the
conclusion that while the joint probabilities such as $\Pr(A,\Bt)$ make sense,
conditional probabilities such as $\Pr(B\vb A)$ do \emph{not} make sense.  And
since $\Pr(B\vb A)$, the probability of $B$ \emph{given} $A$, makes no sense,
this can only be because $A$ is not ``given'', that is, it is not part of
physical reality.

	The preceding sentence is a paraphrase of what Mermin says in the third
to last paragraph of Sec.~IX of \rcite{Mrmn98b}.  He words it a little
differently, making use of the fact that $A$ is, in his rendering of Hardy's
paradox, a property associated with a subsystem, and from this concludes that
there is something suspicious about properties of subsystems.  But since the
analysis given above makes no reference to subsystems, I think Mermin's
reasoning ought to exclude from physical reality any quantum property (subspace,
projector) which might conceivably be a participant in a similar sort of
argument leading to a contradiction.  It seems unfair to limit attention to
properties that refer to subsystems or properties that can serve as conditions
inside conditional probabilities.  In any case it seems likely, given that a
system can be split into subsystems in various ways, etc., that few interesting
properties (0 and $I$ are uninteresting properties) will escape banishment into
nonexistence. Hence it seems to me that, modulo one or two unimportant nuances,
Mermin's ``correlations without correlata'' asserts the existence of any
probability $\Pr(E)$ of the form \eqref{eqn2}, along with the nonexistence of
every $E$.  In probabilistic terms, the probabilities exist, but there is no
sample space or event algebra.

	This is a rather startling solution (assuming it is one) to the problem
of introducing probabilities into quantum mechanics.  Notice that the
contradiction requires the noncommutivity in \eqref{eqn3}, for otherwise
quantum probabilities assigned using \eqref{eqn2} obey all the usual rules of
probability theory.  Therefore what one encounters in Hardy's paradox,
formulated in this way, is the basic \emph{logical} problem pointed out by
Birkhoff and von Neumann, Sec.~\ref{sct2b}, manifested in the fact that the
usual rules for probabilities don't work the way they should.  It is useful to
compare Mermin's solution to this problem with those discussed earlier.
Birkhoff and von Neumann allow all propositions, all projectors, to be part of
a single logical structure, and get rid of contradictions of the sort Mermin
(or Hardy) have discovered by appropriate alterations of the rules of logic.
The histories approach with its single framework rule denies the validity of
certain logical combinations which Birkhoff and von Neumann would allow, and
thereby preserves classical rules of reasoning on appropriately circumscribed
domains of discourse.  Mermin's solution is to
\emph{abolish} the entire logical structure based on subspaces of the Hilbert
space, or at least deny that this has anything to do with the physical world.
I know of no other quantum interpretation which even comes even close to this
in terms of pure philosophical audacity. The Birkhoff and von Neumann scheme of
merely revising propositional logic is rather tame compared with a proposal to
abandon it altogether when dealing with the quantum world!

	I find Mermin's proposal to be at least an implicit endorsement of the
idea, which quantum historians share with Birkhoff and von Neumann, that the
conceptual problems of quantum mechanics are fundamentally \emph{logical}
problems, that is to say, they arise because at least some of the rules of
reasoning with which we are familiar in everyday life break down, do not apply,
or need to be changed in some way when we come to the quantum domain.  There is
a sense in which the great revolutions of modern science consist in calling
into question principles which at least since the time of the ancient Greeks
have been regarded as self-evident, obviously and universally true, and showing
that they are only approximations, or only apply in a limited domain.  The
geocentric universe, the fixity of species, the absolute nature of time, the
validity of Euclid's geometry for describing the world.  Why should the rules
of reasoning be an exception?  My answer to Mermin's question, ``What is
quantum mechanics trying to tell us?'' (the title of \rcite{Mrmn98b}), is that
in the quantum world we have located the limits of validity of the
propositional logic which scientists have applied so successfully in the
classical world. Thus we need to formulate new rules.

	The trouble with Mermin's audacious solution to the logical problem, in
my opinion, is that it goes much too far.  It is throwing out the baby with the
bath water, or sawing off the branch on which you are seated, to express my
dismay in hackneyed phrases that good writers like Mermin would avoid.  Having
gotten rid of the properties or propositions he thought were the source of the
trouble, the result is a conceptual structure that lacks significance.  How
does one attach any meaning to a probability without a referent, to $\Pr(A)$
when $A$ is meaningless?  How does one distinguish it from the
numerically-different probability $\Pr(B)$ of the equally meaningless $B$?

	Another approach to understanding Mermin, which seems initially more
plausible, is that $A$ has a symbolic or mathematical meaning, so that $\Pr(A)$
is the ``probability associated with symbol $A$'', and hence can take on a
different value from the probability $\Pr(B)$ associated with symbol $B$, even
though $A$ and $B$ have nothing to do with the real world.  Theoretical
physicists sometimes use abstract symbols whose physical contents is not clear
--- think of Feynman diagrams in field theory --- but which can be manipulated
according to well-defined rules and used to calculate things, such as cross
sections, which have a closer connection with the real world.  Perhaps for
Mermin the nonexistent referents of his probabilities play this role, as on the
right side of \eqref{eqn2}.  But then what am I to make of the left side?  Can
$\Pr(E)$ be checked experimentally, and if so how?  Normally probabilities are
checked by repeating an experiment many times and looking at the ensemble of
outcomes.  But Mermin rejects ensembles, see Sec.~III of \rcite{Mrmn98b}, as a
way of interpreting his probabilities.  At the end of the day I have to
acknowledge that I am unable to make sense of his objective probabilities, and
I fear that Mermin may be in the same situation; some of his remarks in
Secs.~II and III of \rcite{Mrmn98b} point in that direction.

	A second problem with Mermin's proposal, or maybe it is a different
aspect of the problem just discussed, is its macroscopic or classical or
``correspondence'' limit.  I think it leads, almost inevitably, into a quagmire
called ``consciousness'', and for this reason does not satisfy his desideratum
1 as I interpret it.  Before giving reasons, I should say a few words about my
own position on the proper relationship of quantum physics and consciousness
research.  I think the two need to be firmly disentangled.  This could be done
by producing a correspondence limit for quantum theory which makes sense of the
entities which chemists and biologists are constantly talking about ---
molecules, proteins, cells, what have you --- in quantum terms, one which
explains why such talk is reasonable and shows why what our colleagues believe
to be true is (at least in a suitable approximation) true from a quantum
perspective.  We should be able to say to the neurophysiologists, ``Your way of
talking about things, while not the language of quantum mechanics, is a quite
adequate approximation to what results from a fully quantum-mechanical
description of the sort we physicists know how, in principle, to construct, but
which is extremely awkward and ill-suited to your purposes. You really don't
need it, because your talk about axons, synapses, contact potentials, and the
like makes good physical sense, and is in agreement with our understanding of
the quantum nature of the world.  Stop worrying about Penrose.  Instead, go
solve your problems in the way that seems sensible to you. You have our
blessing.''

	Alas, the Ithaca interpretation cannot pronounce such a blessing,
because ``correlations without correlata'' has the wrong correspondence limit.
There is nothing that allows it to distinguish big macroscopic systems with
enormous Hilbert spaces from microscopic spin systems.  The same sorts of
contradictions can arise for the former as for the latter, and thus,
in the Ithaca interpretation, the probabilities of macroscopic properties make
sense, but the properties themselves do not exist. This immediately raises the
question of why our colleagues (not to mention we ourselves) believe in the
existence of axons and chairs and the like, while supposing that probabilities
are abstract, unphysical things --- exactly opposite to the real state of
affairs according to the Ithaca interpretation.  In this manner the problem of
consciousness, in the form of the mistaken beliefs of a large part of the
scientific community, not to mention the rest of the human race, is immediately
dragged onto center stage in a fashion which I find bizarre.

	Mermin devotes a significant amount of space in \rcite{Mrmn98b} to the
problem of consciousness: it comes up in Secs.~II and X, and all of IV is
devoted to this topic. He does not claim to understand it, and considers it a
significantly more difficult problem than quantum interpretation.  I shall not
try to either summarize or dispute his ideas, as this would require a rather
lengthy essay on a topic for which I possess little expertise.  What he writes
does not indicate that he has managed to cleanly separate consciousness theory
and quantum theory. Instead his strategy, as I understand it, is to plunge
ahead in his effort to understand the quantum mechanical description of the
nonconscious world while leaving inconsistencies between consciousness and his
doctrine of the nonexistence of correlata to be solved at some later date.
	To be sure, his is a new theory, and if no one could propose a new
theory without first solving all the outstanding problems, nothing would get
published.  Theoretical physicists are in the habit of issuing promissory notes
against the day when loose ends will be properly tied down. If I am hesitant to
accept such a note, it is not because I doubt Mermin's honesty.  On the
contrary, it is because I believe him when he says that consciousness is a much
more difficult problem than quantum mechanics.  Which is why the latter needs
to be cleanly separated from the former if we are to make progress, and it
seems to me that the Ithaca interpretation does not accomplish this.

	But can quantum historians do any better?  The answer is a qualified
``Yes''.  First, and this needs no qualification, the ontological ordering of
properties and probabilities in the histories approach is the same as in
classical physics.  Properties correspond to subspaces of the Hilbert space,
the counterparts of subsets of a classical phase space, and this is how
classical physics describes real physical properties.  Probabilities of
properties represent partial information about the system in cases in which
more information is at least potentially available (more on this in
Sec.~\ref{sct4b}).  Thus the Ithaca interpretation's insistence on the
unreality of things most people consider real is absent from the histories
approach, along with the accompanying temptation to start discussing
consciousness.

	Second, and this does need qualification, the histories approach has an
appropriate correspondence limit.  How the classical world emerges from quantum
theory has been studied by Gell-Mann and Hartle,\pocite{GMHr93}
Brun,\pocite{Brn93,Brn94} and Omn\`es.\pocite{Omns99}  While I myself have not
participated in this work, I have tried to summarize the main ideas in Ch.~26
of \rcite{Grff02}. Here is a quick summary of that summary.  Classical physics
emerges when you use a \emph{quasi-classical} framework for quantum theory.  It
employs appropriate projectors, or families of projectors, on (typically rather
large) subspaces chosen so that they represent the properties of
``macroscopic'' (i.e., in comparison with elementary particles) objects used in
everyday discourse, including scientific discourse when it is not focused on
the specifically quantum aspects of matter.  A quasi-classical framework is an
authentic quantum mechanical description of the world, no more and no less than
the unitary or the dragon or the practical frameworks of Sec.~\ref{sct3b}.  A
key point is that a \emph{single} quasi-classical framework suffices for
answering all ``classical'' questions about the world.  (Actually there are
many different, mutually incompatible quasi-classical frameworks which yield
descriptions which are indistinguishable at the macroscopic level; for present
purposes we can simply think of using one of these.) All probabilities which
arise in this framework, and thus all the probabilities needed for classical
physics, can be dealt with using ordinary probability theory, without worrying
about inconsistencies of the type that appear in \eqref{eqn4}.

	The qualification is that actual demonstrations that classical physics
emerges in this fashion are technically quite difficult, as you can see by
looking at Refs.~\citen{GMHr93} and \citen{Omns97b}.  We have a general idea of
how the scheme should work, along with some specific calculations to back it
up.  No nasty surprises have emerged thus far.  But until you have searched the
forest you cannot be sure there are no dangerous animals lurking there, and the
emergence of classical physics from the quantum world covers a lot of territory
still under exploration.

	\subsection{Density matrices as fundamental and objective}
\label{sct4b}
 
	The second pillar of the Ithaca interpretation is that the density
matrix is a fundamental objective property, and provides a complete description
of a single quantum system.  Mermin regards this as a consequence of
desideratum 5, Einstein locality: if $\AS$ and $\BS$ are two systems far enough
apart so that they are not interacting, nothing done to $\BS$ should have any
effect on $\AS$.  It is helpful to consider a specific case: two spin-half
particles $a$ and $b$ in an entangled spin-singlet state $|\psi\rgl$.  The
reader probably knows how this is usually discussed: a measurement of $S_{bz}$
for particle $b$ that yields a value of $+1/2$ results in the wave function
somehow collapsing into a state $|z^-_a\rgl$, $S_{az}=-1/2$, for particle $a$,
while the measurement outcome $S_{bz}=-1/2$ will produce $|z^+_a\rgl$. On the
other hand a measurement of $S_{bx}$ will result in a collapse with
$S_{ax}=-S_{bx}$, and so forth.  Taken literally, this description seems to
imply long-range nonlocal influences that alter (or produce) the state of
particle $a$ depending upon the choice and the outcome of the measurement on
particle $b$.  Einstein did not like such nonlocal influences, and Mermin and I
share his distaste.  But what to do about them?

	One thing you can do is calculate the reduced density matrix
using a partial trace over particle $b$,
\begin{equation}
  \rho_a = \Tr_b([\psi]),
\label{eqn6}
\end{equation}
where $[\psi]$ is short for $|\psi\rgl\lgl\psi|$.  It is then easy to show that
if nothing is perturbing particle $a$ --- no magnetic field, no measurement ---
the choice of which measurement is made on particle $b$ has no effect on
$\rho_a$.  Hence if $\rho_a$ tells us everything that can possibly be said
about the \emph{internal} properties of particle $a$, these cannot be changed
by what is done to particle $b$, and Einstein locality as defined in
desideratum 5 will be satisfied.  By an internal property I mean something like
the value of $S_{az}$, while Mermin means the corresponding probability
distribution.  On the other hand, the \emph{correlation} between particles $a$
and $b$ that tells one that $S_{az}=-S_{bz}$ is \emph{not} an internal property
and cannot be deduced from $\rho_a$.

	Arguments supporting the idea that its reduced density matrix provides
a complete description of the internal properties of a subsystem are given
Sec.~3 of \rcite{Mrmn98}, and are the central focus of \rcite{Mrmn99}.  The
discussion is based on the formula
\begin{equation}
  \rho = \sum_j p_j [\psi_j]
\label{eqn7}
\end{equation}
expressing a density matrix $\rho$ as a sum over terms corresponding to an
``ensemble'' $\{p_j,|\psi_j\rgl\}$, where $\{p_j\}$ is a collection of
probabilities, and $\{|\psi_j\rgl\}$ a collection of normalized, in general
nonorthogonal, pure states.  If $\rho$ is not a pure state there are
(infinitely) many inequivalent ensembles that can be used to represent it in
the form \eqref{eqn7}, and if these ensembles can in some sense be regarded as
physically distinct, then the density matrix by itself is not telling us all
there is to know.  In particular, believers in nonlocal influences will
associate an ensemble $\{|z^+_a\rgl,|z^-_a\rgl\}$, equal probabilities, with
the reduced density matrix $\rho_a$ when $a$ and $b$ are in a spin singlet
state and $S_{bz}$ is measured on particle $b$.  That is, either $S_{az}=+1/2$
or $S_{az}=-1/2$ with the same probability, and we cannot say which since we
are only considering particle $a$ in isolation, not as something correlated
with the outcome of the $S_{bz}$ measurement.  On the other hand, if $S_{bx}$
is measured, the resulting ensemble is $\{|x^+_a\rgl,|x^-_a\rgl\}$.  In this
perspective these ensembles represent distinct states of affairs, despite the
fact that no measurement carried out on $a$ can reveal the difference.  (For
this reason nonlocal influences cannot be used to transmit information, a point
which has received significant attention in the published literature. See, for
example, Sec.~4.6 of \rcite{Rdhd87}.)  Mermin, on the other hand, does not
believe that there are any objective difference between the two ensembles in
this and other analogous cases, and his discussion of the topic in
\rcite{Mrmn99} combines the wit and wisdom for which he is justly famous. I see
no point in trying to repeat or summarize it in my own dull words.  Instead, I
call an intermission during which the reader who has not yet done so can take
time to read and enjoy the dialogue between Yvonne and Zygmund!

	My own assessment of Mermin's argument is that while it represents a
good first step in defending Einstein locality, one can do a lot better, but in
order to do a lot better one needs to \emph{abandon} the idea that the density
matrix of a subsystem is an objective and irreducible description, and instead
treat it as a quantum analog of a classical probability distribution, something
you use when the information available to you is less than what is, in
principle, potentially available.  I shall (of course) be using histories to
discuss the situation, and for this purpose it is helpful to allow for a time
dependent $|\psi(t)\rgl$ obtained by integrating Schr\"odinger's equation
starting with an initial entangled state $|\psi(0)\rgl$ at $t=0$.  
The time dependence comes from a Hamiltonian in which there is no interaction
between the two particles, but one or both of them could be in a magnetic
field. For most of the discussion I will assume zero magnetic field, so that
$|\psi(t)\rgl$ is the same as the initial  $|\psi(0)\rgl$.

	The probability that $S_{az}=+1/2$ at time $t$ can be written as
\begin{equation}
  \Pr(S_{az}=+1/2) = \lgl\psi(t)|\blp [z^+_a]\ot I\brp |\psi(t)\rgl 
  =\Tr_a\blp \rho_a(t)[z^+_a]\brp, 
\label{eqn8}
\end{equation}
where $\rho_a(t)$ is defined by the
time-dependent version of \eqref{eqn6}.  The event algebra on which this
probability is defined must include $[z^+_a]$
and its negation $[z^-_a]$, and the coarsest (smallest) sample space
for which this is true consists of these two projectors.
This sample space does \emph{not} contain $[\psi(t)]$, and it is inconceivable
that it could somehow be extended to include $[\psi(t)]$, because $[\psi(t)]$
does not commute with $[z^+_a]$ or $[z^-_a]$: the projector onto any entangled
state of two spin-half particles does not commute with that for any nontrivial
property of either spin.  Hence it is meaningless to think of the two-spin
system as somehow being in the entangled state $|\psi(t)\rgl$ when at that $t$
one uses $|\psi(t)\rgl$ to compute the probability \eqref{eqn8}.  The careless
habit of physicists in thinking that one can ascribe properties to separate
spins while believing that the combination is in an entangled state is, from
the perspective of the quantum historian, just that --- a careless habit, which
can and does get people into trouble.

	But if $|\psi(t)\rgl$ is not a physical property, how can it appear in
\eqref{eqn8}?  In this context it functions not as a physical property or
quantum beable, but as a mathematical device to assist calculation, a
\emph{pre-probability} in the notation of Sec.~9.4 of \rcite{Grff02}, to which
I refer the reader for a detailed discussion.  If $|\psi(t)\rgl$ in
\eqref{eqn8} is merely a device for calculation and not a physical property,
the same must be true of $\rho_a(t)$.  (Note that $\rho_a(t)$ is not really
needed, since anything one can calculate from it can be obtained just as well
from $|\psi(t)\rgl$, and one can use $|\psi(t)\rgl$ to compute other things,
such as the correlation between the two particles, which can't be obtained from
$\rho_a(t)$.)  Thus $\rho_a(t)$ is a pre-probability and cannot be more real
than a probability, so in the histories approach it is not a beable.  Note once
again how real vs.\ imaginary gets reversed when one goes from Ithaca to
histories.

	Unlike Mermin, the historian claims that a more refined, more precise,
more informative description of particle $a$ is available, at least in
principle, than that provided by $\rho_a$.  For the spin singlet state,
$\rho_a$ assigns probabilities of 1/2 to each of the two possible values of
$S_{az}$.  This is very little information, and corresponds to not knowing the
outcome of a coin toss.  How can we do better?  By \emph{measuring} $S_{az}$
using a Stern-Gerlach apparatus, and seeing whether its value is $+1/2$ or
$-1/2$.  Alas, you say, the measurement will perturb the particle and the
outcome will not tell us whether $S_{az}$ was $+1/2$ or $-1/2$ \emph{before}
the measurement.  You will indeed learn nothing if you insist on using the
dragon framework, Sec.~\ref{sct3b}, that you learned from your textbook.
Instead use an appropriate framework, the same thing experimental physicists do
when they design real apparatus to measure real properties.  They are not
fools!

	Having reduced Mermin's irreducible description, can we still defend
Einstein locality?  Indeed we can, with a much more robust and compelling
argument than Mermin's.  For one can show --- see Ch.~23 of \rcite{Grff02} for
the detailed justification --- that if at time $t_1$ particle $a$ has, for
example $S_{az}=+1/2$, and at $t_2 > t_1$ some spin component (it doesn't
matter which one) of particle $b$ is measured, then at any time $t_3 > t_1$,
including $t_3 > t_2$, it is still the case that $S_{az}=+1/2$.  (This assumes
that particle $a$ is not in a magnetic field.) You can replace $S_{az}=+1/2$
with $S_{aw}=+1/2$, where $w$ is any direction in space, and result is the
same: the distant measurement leaves it unchanged.  In brief, the historian can
carry out an explicit calculation that demonstrates Einstein locality. And as a
corollary provides a very simple proof that nonlocal influences cannot
transport information: they don't exist!

	This argument is not available to Mermin, who
thinks $S_{az}=+1/2$ is meaningless. However, a modified form can be
constructed for  someone who only believes in correlations. The fact that 
nothing happens to $S_{az}$ when $b$ is measured at $t_2$ can be expressed 
using joint probabilities 
\begin{equation}
\begin{split}
  \Pr([z^+_a], t_3;\, [z^-_a], t_1) &=0,
\\
  \Pr([z^-_a], t_3;\, [z^+_a], t_1) &=0.
\end{split}
\label{eqn9}
\end{equation}
which are easily calculated using histories, see Ch.~23 of \rcite{Grff02} for
the general strategy. I myself find it a bit clearer to write
\begin{equation}
\begin{split}
  \Pr([z^+_a], t_3 \vb\, [z^+_a], t_1) &=1,
\\
  \Pr([z^-_a], t_3 \vb\, [z^-_a], t_1) &=1,
\end{split}
\label{eqn10}
\end{equation}
but \eqref{eqn9} can be used by someone suspicious of conditionals.

	However, there is no way to derive either \eqref{eqn9} or \eqref{eqn10}
from $|\psi(t)\rgl$ or from the reduced density matrix $\rho_a(t)$.  This
seems to have escaped people who suppose that the ``wave function of
the universe'',  thought of as a function of time satisfying
Schr\"odinger's equation, somehow provides a complete description of the
time dependence of a quantum system.  One way to see why this is not so is
to use the classical analogy of a Brownian particle that starts at the origin
at  $t=0$.  The usual simplified
theory leads to a probability distribution density 
\begin{equation}
  \rho(\rb,t) = (4\pi D t)^{-3/2} \exp[ -r^2/4D t],
\label{eqn11}
\end{equation}
for the Brownian particle to be located at a point $\rb$ at time $t$, where $r$
is the magnitude of $\rb$ and $D$ is the corresponding diffusion constant.
	This $\rho(\rb,t)$ provides a useful probabilistic description,
something that can be checked against experiments.  But even on a probabilistic
level there is more that can be said.  Suppose that at time $t_1=5$ seconds the
particle is at $\rb_1$.  Then at $t=5 +\Dl t$ seconds the probability of its
location conditional on this information will not be given by \eqref{eqn11},
but by a different expression involving $\rb-\rb_1$ and $\Dl t$.  The point is
that you cannot use the \emph{single} time-dependent expression \eqref{eqn11}
to describe what is going on, for it lacks information about
\emph{correlations} of positions at different times.  
Clearly \eqref{eqn11} does not describe the motion of an \emph{individual}
Brownian particle. It is more appropriately thought of as describing an
ensemble: a collection of many Brownian particles, or the results of 
repeating an experiment on one Brownian particle many times. 

	In the same way, $\rho_a(t)$ in our quantum example is missing the sort
of information needed to derive \eqref{eqn9} or \eqref{eqn10}.  That $S_{az}$
has a probability of 1/2 of being $+$ at both $t_1$ and $t_3$ is consistent
with its being the same at both times, \eqref{eqn9}, but also with its having
reversed its sign at some point during the intervening time interval, say at
$t_2$ when the measurement was carried out on particle $b$.  There is nothing
wrong with $\rho_a(t)$, but it simply is not an appropriate tool for
calculating correlations of properties at different times.  The same comments
apply to $|\psi(t)\rgl$.  Quantum histories, on the other hand, are an 
appropriate tool, and for this reason satisfy
Mermin's desideratum 3, which says we should be able to \emph{describe} the
behavior of individual systems even in cases in which the theory cannot provide
deterministic \emph{predictions}.  The Ithaca interpretation, at least as long
as it relies on $\rho_a(t)$ or $|\psi(t)\rgl$, cannot obtain \eqref{eqn9},
and is limited to what I think is best thought of, using the analogy between
$\rho_a(t)$ and $\rho(\rb,t)$, an ensemble rather than a single
system description.

	\section{Reality and Its Representations}
\label{sct5}

	Mermin has a brief discussion of quantum histories in Sec.~XI of
\rcite{Mrmn98b}.  In a footnote he remarks that the term \emph{incompatible} as
used in the histories approach has something in common with Bohr's
\emph{complementary}, and he thinks it would be better if the latter were used
in place of the former. My response is that while there is a relationship
between the concepts, there is also a risk of confusion, a point to which I
shall return below.  Mermin goes on to say that the histories approach
liberates Bohr's complementarity from the context of mutually exclusive
experimental arrangements, and then adds:

\begin{quote}

	The price that one pays for this liberation is that the paradoxical
quality of complementarity is stripped of the protective covering furnished by
Bohr's talk of mutually exclusive experimental arrangements, and laid bare as a
vision of a single reality about which one can reason in a variety of mutually
exclusive ways, provided one takes care not to mix them up.  Reality is, as it
were, replaced by a set of complementary representations, each including a
subset of the correlations and their accompanying correlata.  In the consistent
histories interpretation it is rather as if the representations have physical
reality but the representata do not.

\end{quote}

	The last sentence is at first glance rather perplexing, as the
historian's motto of ``no probabilities without a sample space'' translates
into ``no correlations without correlata,'' which would seem to correspond to
``no representations without representata,'' the exact opposite of what Mermin
asserts.  I am indebted to him for the following clarification:\pocite{note1}
``representations'' are frameworks in histories terminology, and the denial of
physical reality to the representata was intended to refer to the impossibility
of combining sample spaces of separate incompatible frameworks into a single
sample space.  Even with this clarification I find myself in almost complete
disagreement with Mermin's statement, and since the issues involved are rather
central to a correct interpretation of the histories approach, I shall attempt
to briefly state my own position on the relationship of frameworks to the
physical reality they in some sense represent.

	To begin with, only a very naive realism would assert that the
mathematical constructs used in theoretical physics are identical with the
physical world which many of us believe is ``out there,'' independent of our
attempts to describe it, and to a great extent immune to our efforts to change
it.  Instead, our theories are, at best, abstract mathematical representations
of that reality.  Electrons do not solve Schr\"odinger's equation, protons do
not live in Hilbert space, and even in the pre-quantum era planets did not move
on orbits in phase space.  The relationship of our mathematical and conceptual
structures to the world ``out there'' is a subtle and difficult philosophical
problem.  My attitude, perhaps better my faith, is that reality is
\emph{something like} our best scientific models of the world.  In particular,
a Hilbert space of wave packets and spins, while it is only an abstract model,
reflects reality better than does a classical phase space.

	The phase space is not so much wrong as outdated, and it served very
well as a representation of physical reality into the early 20th century.
Since a point in phase space comes as close as possible to representing or
describing a classical system as we think it ``really is,'' it seems sensible
to assign the same task to a ray in Hilbert space, the closest quantum analog
of a point in phase space.  The ray is, in this sense, a mathematical
representation of what I believe, but cannot prove, is the reality ``out
there'' in an objective world.  This connection between the quantum Hilbert
space and the real world, which seems to me central to the physical
interpretation of quantum histories, is what is being denied by the Ithaca
interpretation's ``correlations without correlata.''

	Let us return to the problem which seems behind Mermin's denial that
the histories approach provides a consistent representation of reality: the
fact that incompatible sample spaces cannot be combined.  Part of the
difficulty may be that ``incompatible'' used in this way is a technical quantum
term referring to something with no good classical analog. In particular, it
means something very different from ``mutually exclusive'' as applied to
distinct elements of a classical (or quantum) sample space.  Various critics
have confused these two terms and thereby seriously misunderstood what the
histories approach is all about, and this is one reason I do not think it is a
good idea to replace ``incompatible'' with Bohr's ``complementary'', at least
insofar as the latter is associated (Bohr was hardly a model of clarity) with
mutually exclusive pieces of apparatus.  The following comments will, I hope,
help clarify the difference; lengthier discussions will be found elsewhere
(\rcite{Grff98} and Ch. 27 of \rcite{Grff02}).

	For a spin-half particle, $S_x$ and $S_z$ are \emph{incompatible
observables}, and the projectors $[z^+]$ and $[x^+]$ represent
\emph{incompatible properties}, since $[z^+][x^+]$ is not zero and not equal to
$[x^+][z^+]$.  In order to \emph{measure} $S_x$ for a particle traveling along
the $y$ axis one needs to rotate the Stern-Gerlach apparatus through $90\dg$
from the orientation needed to measure $S_z$, and these two are \emph{mutually
exclusive} experimental arrangements, as are any two macroscopically-distinct
states of affairs when viewed quantum mechanically: the product of the
projectors is zero.  Note that $[z^+]$ and $[x^+]$ do not and cannot belong to
a single quantum framework, whereas the two orientations of the apparatus can
and do belong to a single quasi-classical framework.  For this reason,
measurements, while they can supply useful insights, are not a good, or at
least an adequate way to try and understand quantum incompatibility.  Instead,
one has to build up one's intuition on the basis of microscopic quantum
examples analyzed in a consistent way that is not dependent upon a notion of
macroscopic measurement.  To this end I recommend toy models of the sort used
extensively in \rcite{Grff02}.

	Even though there is no really good classical analog of quantum
incompatibility --- which is why it is so hard to get it straight --- judicious
use of classical models, such as the following, can sometimes help.  Imagine a
spinning object with angular momentum $(L_x,L_y,L_z)$, and suppose that for
some reason I describe it by giving the value of $L_z$ and not that of the
other two components.  Such an impoverished description constitutes the $L_z$
framework, and $L_x$ and $L_y$ frameworks are defined in the same way.
	Using the $L_z$ framework doesn't mean one cannot use the $L_x$
framework, and whether or not some $L_z$ description is true has no bearing on
whether an $L_x$ description is true.  Clearly no ``law of nature'' tells one
that one of these frameworks is ``correct'' and the others are ``incorrect.''
Instead, if it is the $z$ component of angular momentum that interests me, I
will need to use the $L_z$ framework, because in the $L_x$ and $L_y$ frameworks
$L_z$ is (obviously) not defined.  Also I can, if I wish, use the $L_z$
framework up to some point in time, and then switch over to the $L_x$ framework
for later times without altering the angular momentum of the body; all I have
done is switch my attention from one aspect to another.

	Now consider a quantum spin-half particle with its $S_x$, $S_y$, and
$S_z$ frameworks.  \emph{Every statement in the preceding paragraph is still
correct if $L$ is replaced by $S$!}  Go back and read it over again with this
in mind.  The difference between classical and quantum is that you \emph{cannot
combine} two incompatible quantum descriptions, so specifying values of both
$S_x$ and $S_z$ for the same particle at the same time is meaningless, whereas
the classical descriptions are always compatible and therefore always
combinable.  This is why the discussion of classical angular momentum in the
previous paragraph seems so unnatural, even though it is correct: in classical
physics one never has any reason to talk about $L_z$ while not talking about
$L_x$ and $L_y$.  The situation in quantum physics is different, but it is
nonetheless misleading to think of the $S_x$ and $S_z$ descriptions as mutually
exclusive, as if one being true implies that the other is false.  Which
framework is to be used depends on the sort of description the theoretical
physicist wants to construct.  He is of course perfectly free to construct all
sorts of incompatible descriptions and contemplate them in any way he finds
helpful in better understanding the quantum world.  However, he will be badly
mistaken if he thinks he can combine two of them into a single description that
somehow corresponds to the reality of what is going on in his colleague's
laboratory.
	This may be something of what Bohr had in mind when he spoke of
``complementarity,'' and I would like to believe that it is. But if so,
anchoring the concept firmly in the mathematical structure of Hilbert space
rather than tying it to possible arrangements of macroscopic apparatus is a
necessary step in producing a tool of real utility to the quantum physicist,
something that can go beyond the usual arm-waving appeal to the uncertainty
principle and be used to disentangle something like Hardy's paradox. 

	A particularly important framework for understanding the histories
approach to quantum reality is the quasi-classical framework mentioned earlier,
the one appropriate for describing the everyday world.  As already noted in
Sec.~\ref{sct4a}, this framework provides what is in principle (it would,
obviously, be hideous to use it in practice) a genuine quantum-mechanical
description of how things develop in time, using histories based on
projectors representing actual properties of the world.  The dynamics is
stochastic, and only one of the histories to which quantum mechanics assigns
probabilities will correspond to what actually happens.  But what actually
happens actually happens, and in this respect the histories approach provides a
robust realism in marked contrast to the Ithaca interpretation.  What about MQS
states, Schr\"odinger's cat?  There are certainly frameworks which contain such
things, and no law (in particular, no law of nature) prevents physicists from
using them.  Think of such a framework as describing the ``$S_x$'' aspect
of the world in a situation in which the quasi-classical framework describes
the ``$S_z$'' aspect.  These descriptions are \emph{not} mutually exclusive in
the sense that if one is correct the other must be false.  Think again of
classical $L_z$ and $L_x$. But you cannot combine them, for attempting to do so
produces something which doesn't live in the Hilbert space.

	Returning to the topic of physical reality as viewed quantum
mechanically, it seems to me rather natural to allow it to include everything
that can be described by a quasi-classical framework, with the obvious proviso
appropriate for a stochastic theory: only one of the histories actually occurs.
This way everything that is part of reality in macroscopic classical physics
has a quantum counterpart.  Including other quantum frameworks on an equal
footing as possible or potential descriptions of physical reality is then a
natural idea, and there seems to be nothing wrong with it so long as one keeps
in mind that quantum reality understood in this way is such that its diverse
incompatible aspects cannot be represented by a single picture of the world in
the manner we are accustomed to in classical physics.  To be sure, one may also
take the position that unless the frameworks (representations) satisfy
classical combination rules, what they represent cannot count as part of
genuine physical reality.  If this is Mermin's position, I disagree with him,
for I believe that the most interesting developments of modern science are
precisely those that have caused significant changes in our view of reality,
and quantum mechanics belongs on that list.

	\section{Conclusion}
\label{sct6}

	The Ithaca interpretation and the histories approach share a common
vision of what constitutes a good interpretation of quantum mechanics.  As
indicated in Sec.~\ref{sct1}, I agree wholeheartedly with the first five of
Mermin's six desiderata, and for reasons which I have alluded to from time to
time in this paper, I believe the histories approach, as summarized in
Sec.~\ref{sct3a} satisfies all five. On the other hand, it seems to me that the
Ithaca interpretation is deficient in two respects.
	
	The first and most serious problem is connected with the first
desideratum: The theory should describe an objective reality independent of
observers and their knowledge.  If words are to have their usual meaning,
``reality'' must somehow be connected with what most people, including most
quantum physicists, think constitutes the real world, with its chairs and
photomultiplier tubes and neurons and the like.  And most of us think that the
\emph{actual} outcome of a coin flip --- it came up tails on this particular
occasion --- is \emph{more} real than the probability of 1/2 assigned to it
before the experiment took place.  But Mermin claims, if I have not
misunderstood him, that in the quantum world probabilities are real and their
referents are not; my difficulties with this are explained in Sec.~\ref{sct4a}.

	To be sure, the quantum world may be sufficiently different from the
world of everyday experience that words cannot have their usual meaning, and I
think this must to some extent be the case.  All of the great revolutions of
modern science have required some revision of the meaning of words. Think of
how our notion of ``time'' was altered by relativity theory.  It would be
disappointing if quantum mechanics, which is surely much more revolutionary
than relativity, did not require similar changes.  In Sec.~\ref{sct2a} I
explain why the use of Hilbert space rather than phase space mathematics in
quantum theory raises crucial issues about how we reason about the world, and
in Secs.~\ref{sct2b} and \ref{sct3} some proposals for dealing with this.  I
can hardly criticize innovation when the histories approach that I advocate
requires a major revision of ``classical'' ways of thinking when dealing with
the quantum world.

	However, quantum logic, quantum histories, and Copenhagen or textbook
quantum mechanics (at least when viewed from a histories perspective) have in
common the fact that in the correspondence limit of macroscopic systems their
vision of reality corresponds, pretty much, to what we believe about the
ordinary world.  The Ithaca interpretation does not.  This does \emph{not} by
itself mean that it is wrong: remember that Galileo's opponents thought it
obvious that the earth was at rest, whereas we have come to accept the fact
that it is, instead, in rapid motion.  But it does mean that a lot more work
needs to be done to make sense out of the Ithaca interpretation, and if a
theory of consciousness is needed in order to explain why the commonsense view
of reality is wrong, I am rather pessimistic about the project.  At least we
need to explore the possibility that quantum mysteries can be satisfactorily
solved in a less radical manner.

	The other deficiency I see in the Ithaca interpretation has to do with
desideratum 3: The theory should describe individual systems --- not just
ensembles.  Here the issues are more subtle.  The histories approach is
capable, Sec.~\ref{sct4b}, of providing a more detailed description of the
properties of one of the two particles prepared in an entangled state than is
possible using a reduced density matrix, contrary to Mermin's claim that the
latter is objective and irreducible.  To use an analogy, the historian is
willing to say that the outcome of \emph{this particular} coin toss was $T$,
whereas Mermin can only assign equal probabilities to $H$ and $T$. I myself
have no problem with assigning a probability to a particular event, such as
whether or not it will snow in Ithaca on December 25 of 2010. But it seems to
me that such an assignment does not really describe the event, and by analogy
the reduced density matrix does not describe an individual system, even though
it may supply us with some useful information. And to the extent that a
reduced density matrix $\rho_a(t)$ is the quantum analog of the Brownian
particle probability distribution in \eqref{eqn11}, it seems appropriate to
think of it as describing an ensemble rather than an individual system.  

	I would agree that the Ithaca interpretation satisfies the other
desiderata: it describes small systems, measurements do not play a fundamental
role, and it preserves Einstein locality.  In the case of Einstein
locality (Sec.~\ref{sct4b}) it seems to me that the histories approach achieves
sharper results, but at least Mermin and I agree on the absence of those
mysterious nonlocal influences.

	The Ithaca interpretation is much younger than the histories approach,
and Mermin is quite frank in acknowledging that his notion of objective
probability is tentative, and that there are a number of other unresolved
problems and loose ends; see in particular Sec.~XII of \rcite{Mrmn98b}.  These
will need to be tied down a bit better before one can say whether or not this
approach is really effective in taming quantum mysteries.  In the meantime the
histories interpretation shows that the vision represented by his first five
desiderata is not just wishful thinking on Mermin's part.  There is already one
consistent interpretation of quantum mechanics that fits the bill, and it will
be interesting to see if this can also be done in a completely different way.
It is high time to clean up the conceptual mess which has produced such
headaches for physicists (I am particularly concerned about beginning
students), and served as a breeding ground for all manner of crazy
philosophical oddities.  Even if Mermin and I do not agree on how to accomplish
this, we seem to share a remarkably similar vision of what the result ought to
look like when the task is complete.

	\section*{Acknowledgments}

	It is a pleasure to acknowledge helpful conversations and
correspondence with David Mermin in my efforts to understand the Ithaca
interpretation, and in writing the present article.  My research has been
supported by the National Science Foundation Grants 9900755 and 0139974.


\end{document}